\documentclass[prb,aps,twocolumn,epsfig]{revtex4}
\usepackage{graphicx}
\usepackage{dcolumn}
\usepackage{bm}

\begin{document}
\title {Signature of Martensite transformation on conductivity noise in thin films of NiTi shape memory alloys}

\author{Chandni. U\footnote[1]{electronic mail:chandni@physics.iisc.ernet.in}
} \author{Arindam Ghosh\footnote[2]{electronic
mail:arindam@physics.iisc.ernet.in}}
\address{Department of Physics, Indian Institute of Science, Bangalore 560 012, India}
\author{H. S. Vijaya and S. Mohan }
\address{Department of Instrumentation, Indian Institute of Science, Bangalore 560 012, India}

\begin{abstract}
Slow time-dependent fluctuations, or noise, in the electrical
resistance of dc magnetron sputtered thin films of Nickel Titanium
shape memory alloys have been measured. Even in equilibrium, the
noise was several orders of magnitude larger than that of simple
diffusive metallic films, and was found to be non-monotonic around
the martensitic transformation regime. The results are discussed
in terms of dynamics of structural defects, which also lay
foundation to a new noise-based characterization scheme of
martensite transformation.
\end{abstract}


\maketitle

Growing interest in thin films of Nickel-Titanium (NiTi) shape
memory alloys have been driven by many applications such as
micropumps, microactuators for micro-valves and micro-positioners,
etc.~\cite{otsuka,otsuka2}. The thermoelastic martensite
transformation causes shape memory effect as well as
pseudo-elasticity, which makes these systems excellent actuation
elements in microelectromechanical designs. However, in the thin
film form, the diffusionless structural transformation from the
high temperature (cubic B2: CsCl) austenite phase to low
temperature (monoclinic B19/B19$^\prime$) martensite
phase~\cite{huang} may be influenced by external parameters
including the substrate, finite grain size, surface defects
etc.~\cite{lee}, which often pose difficulty in probing the
intrinsic structural dynamics in an unambiguous manner. The
information from differential scanning calorimetry (DSC), Xray
diffraction (XRD), scanning thermal microscopy (SThM), acoustic
emission measurements etc.~\cite{otsuka2,reche}, are often
limited, and it is desirable to have a more microscopically
intuitive understanding of the martensite dynamics and
corresponding temperature scales.

Here, we employed a new technique to characterize the martensite
phase in thin NiTi films. The technique is based on measuring the
time-dependent low-frequency fluctuations, or noise, in electrical
resistivity ($\rho$) at fixed temperatures ($T$) at various stages
of the austenite-to-martensite phase transition. Since electronic
transport properties are directly affected at the martensite
transition, for example, by the change in structural and
scattering properties, or that in the density of states at the Ti
sites due to the lowering in energy of the B19/B19$^\prime$
structure, the time-averaged $\rho$ vs. $T$ has long been used as
an empirical indicator of martensite transition~\cite{Nam}.
Equilibrium resistance noise arises from the coupling of electron
transport to slow variations in the configuration of structural
disorder including migration/rotation of defect or atomic species,
scattering off slow fluctuators or two-level systems etc. While
noise at stress-induced reorientation in the martensite phase of
free-standing NiTi wires has been measured before~\cite{arindam},
no systematic investigations exist on the signature of martensite
formation on equilibrium $T$-dependence of noise. Such a study
would probe the influence of grains and grain boundaries on the
thermoelastic properties of the NiTi films, and provide
independent and microscopically intuitive estimates of the
temperature scales of martensite transformation that may be
relevant to growing applications of NiTi thin films.

The samples were prepared by dc magnetron sputtering of a mosaic
target which consists of patterned Titanium disk of $76$ mm
diameter and $0.8$ mm thickness laminated over a circular Nickel
disk of $76$ mm diameter and $1.6$ mm thickness. Si $(100)$
cleaned with dilute HCl and methanol were used as substrates. Two
samples (S1 and S2) were deposited under similar conditions (Ar
pressure of $2 \times 10^{-3}$mbar) with thicknesses $0.9\mu$m and
$1\mu$m respectively, but annealed at dissimilar temperatures
($480^\circ$~C and $400^\circ$~C respectively) that resulted in
different levels of surface defects. The samples were
characterized using XRD and scanning tunnelling microscopy (STM),
which indicated coexistence of martensite and austenite phases at
room temperature, with grain size $\sim 160$nm. STM studies also
indicated higher rms surface roughness in S2 than in S1. To
measure noise, we have used a $5$-probe ac measurement
technique~\cite{scofield,arindam2} with a resolution in the noise
power spectral density ($S_V$) of $\approx
1\times10^{-20}$~V$^2$/Hz. The background noise primarily consists
of the Johnson's noise, and atleast two orders of magnitude lower
than the device noise at all $T$.

\begin{figure}
\begin{center}
\includegraphics[width=8.5cm,height=6cm]{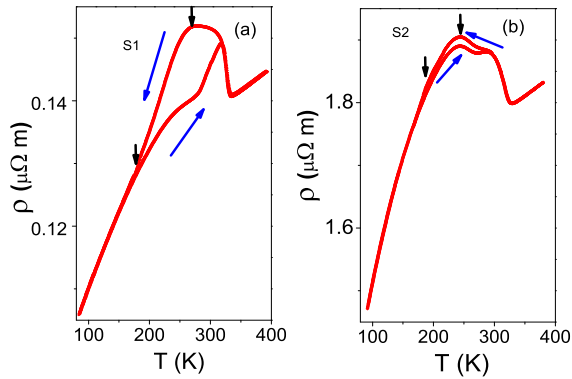}
\end{center}
\vspace{-0.7cm} \caption{Resistivity vs Temperature curves for (a) S1 and (b) S2. The arrows indicate the heating and cooling directions.
$M_{s}$ and $M_{f}$ are also indicated by arrows.} \label{figure1}
\end{figure}

Although both samples show qualitatively similar metallic behavior
and characteristic hysteresis, resistivity ($\rho$) data in Fig.~1
indicates $\rho$ in S2 to be more than one order of magnitude
higher than that of S1 confirming the presence of excess disorder
scattering. In both cases the rise in $\rho$ around 320~K on the
cooling cycle implies formation of the R-phase on going from the
cubic B2 (austenite) to the monoclinic B19$^\prime$ (martensite)
crystal structure~\cite{otsuka}. We note that (1) in spite of
large difference in $\rho$, the absolute magnitude of the
hysteresis in $\rho$ is similar for both samples. This confirms
that the excess disorder in S2 only enhances the disorder density
and the scattering rate, without substantially affecting the
structural morphology of the system. (2) Secondly, the martensite
start and finish temperatures can also be identified in the
conventional manner~\cite{otsuka2}, which turn out to be $M_s
\approx 270K$ and $M_f \approx 180K$ for S1 and $M_s \approx 245K$
and $M_f \approx 190K$ for S2.

In order to focus on the equilibrium dynamics in the films, the
noise was measured at fixed $T$ (stabilized to an accuracy of
3~ppm) after a waiting period of 1800 - 2000 sec at each $T$.
Fig.~2 shows typical time-dependent fluctuations in $\rho$ in
three different regimes in S1: (a) the $T = 360$~K data represents
fluctuations deep into the austenite phase, (b) at $T = 260$~K,
where the $\rho - T$ data suggests onset of martensite formation,
and (c) at $T = 110$~K, where a substantial drop in the noise
magnitude is seen deep into the martensite phase. The
corresponding power spectral densities $S_\rho$ of noise are shown
in Fig.~2b. Apart from the decrease in magnitude from austenite to
martensite we find $S_\rho \sim 1/f^\alpha$, where the frequency
exponent $\alpha \sim 1 - 1.2$, and resembles the conventional
$1/f$-noise observed in a wide range of disordered metallic thin
films~\cite{fleetwood,weissman}. We point out that the behavior of
$S_\rho$ in stress-driven resistance fluctuations indicated
$\alpha$ as large as $\sim 3$ which was attributed to burst-like
dynamics~\cite{arindam}.

The $T$ variation of equilibrium noise magnitude is shown for both
samples in Fig.~3a. For comparison, the measured $S_\rho$ is
normalized to the conventional Hooge parameter $\gamma_H$ defined
as,

\begin{displaymath}
S_{\rho}(f)=\frac{\gamma_H \rho^{2}}{n\Omega f^{\alpha}}
\end{displaymath}

\noindent where $n$, $\Omega$ and $\rho$ are number density of
atoms, sample volume, and the resistivity of the sample,
respectively. Note that the observed magnitude of $\gamma_H$ is
$\sim 9$ orders of magnitude higher than for typical diffusive
metallic films~\cite{Dutta,scofield2}. Fig.~3a shows both heating
and cooling cycles. Note that $\gamma_H$ in S$2$ is nearly two
orders of magnitude larger than that of S$1$, indicating the
pivotal role of defects in producing the fluctuations~\cite{pelz,
kogan}.

If noise originates from defect dynamics, the $T$-dependence of
$\gamma_H$ is expected to be directly proportional to the density
of mobile defects~\cite{pelz}. Here, $\gamma_H$ shows a nearly
activated behavior with T deep inside the martensite and austenite
regimes (Fig.~3a and 3b), with activation energies of $\approx
0.17$~eV and $0.28$eV in the martensite and austenite phases
respectively. The typical energy scales of defect (void or
interstitials) creation in NiTi is 1-2eV~\cite{russell}, however,
presence of grains and grain boundaries can modify this energy
scale substantially. Moreover, contributions from additional
mechanisms, such as interference effects~\cite{arindam3}, cause
noise to deviate from pure activated form at very low $T$ ($\ll
M_{f}$).

\begin{figure}
\begin{center}
\includegraphics[width=8.5cm,height=6cm]{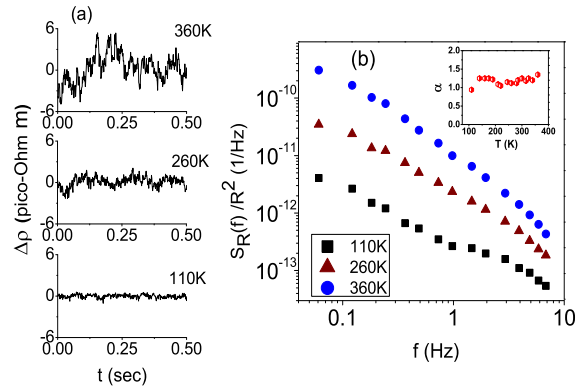}
\end{center}
\vspace{-0.7cm} \caption{(a)Resistivity Fluctuations for heating, at temperatures $110$ K, $260$ K and $360$ K as a function of time. (b)
Normalised Power Spectral Density as a function of frequency, for the three temperatures. Note that the behavior is 1/f$^{\alpha}$ ,
{$\alpha$}{$\sim$}1-1.2 at all temperatures. $\alpha$ (at $1Hz$) vs T is plotted in the inset. } \label{figure1}
\end{figure}

A striking feature in the behavior of $\gamma_H$ is the weak or
slightly decreasing $T$-dependence over $\sim 200 - 300$~K,
irrespective of heating or cooling cycle. From the $\rho - T$ data
(Fig.~1) we find this region to appear between $M_s$ and $M_f$ for
both samples. A plausible mechanism leading to the slow variation
of $\gamma_H$ between $M_s$ and $M_f$ may involve a two-phase
coexistence during martensite formation. Assuming individual Hooge
parameters to be $\gamma_H^A$ and $\gamma_H^M$ in the austenite
and martensite regions respectively, the net Hooge parameter at an
intermediate $T$, $M_s > T > M_f$, can then be phenomenologically
expressed as $\gamma_H = f\gamma_H^M + (1-f)\gamma_H^A$, where $f$
is the fraction of martensite region that changes from $0
\rightarrow 1$ as $T$ varies from $M_s \rightarrow M_f$, resulting
in a weakly varying $\gamma_H$ as observed in the transition
regime. XRD data also confirms the coexistence of the martensite
and austenite at room temperature in sample S$2$, although the
signature of austenite was weak in S$1$. Alternatively, a partial
annealing of defects~\cite{pelz}, may also occur as the crystal
reorients between $M_s$ and $M_f$ resulting in a weakly decreasing
$\gamma_H$ with increasing $T$ in the heating cycle. Similar
behavior in the cooling cycle possibly indicates generation of new
defects, as the crystals transform from high symmetry austenite to
low symmetry martensite phase. Nevertheless, the sharp changes in
the $T$-dependence of noise at the beginning and end of the
martensite allows independent estimates of $M_s$ and $M_f$.

\begin{figure}
\begin{center}
\includegraphics[width=8.5cm,height=6cm]{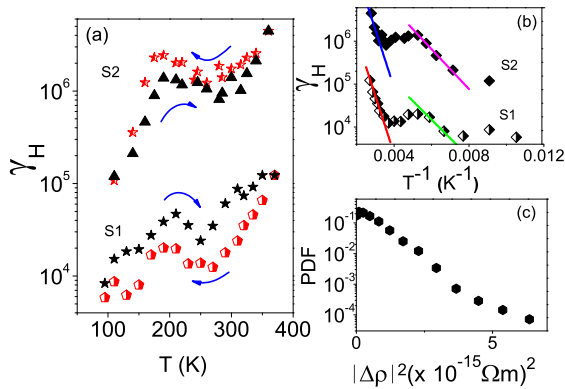}
\end{center}
\vspace{-0.7cm}\caption{(a) Hooge Parameter $\gamma_{H}$ vs T for both the samples. Note that the regions at the two ends are linear and the
middle portion is flattened out. (b) The activated behaviour of the two phases for samples $S1$ and $S2$ during cooling. (c) Normalised
Probability Density Function (PDF) vs. $\Delta\rho^{2}$ for S$1$.} \label{figure1}
\end{figure}

Another crucial aspect of the data shown in Fig.~3a is the
pronounced hysteresis in the noise data between heating and
cooling cycles, where the range of hysteresis in $T$ largely
exceeds that observed in the simple $\rho-T$ measurements
(Fig.~1). This is particularly surprising considering the
equilibrium nature of the noise measurement. We believe its origin
lies in substantial modification of the transformation temperature
scales along the grain boundaries in NiTi thin films. The
interaction of the martensitic plates with grain boundaries would
not only be sensitive to the grain size, but also the physical
structure of the grains, leading to a wide distribution of $M_s$,
$M_f$, $A_s$ and $A_f$ along the grain
boundaries~\cite{huang,gil}. The sensitivity of noise to local
microstructures and corresponding energy scales can be readily
understood, if we assume noise to be primarily due to diffusion of
defects along the grain boundaries. Time-averaged resistivity
$\rho$, on the other hand, is determined by electron scattering in
the bulk of the grains as well, and less sensitive to the nature
of microstructure at the grain boundaries.

The intriguingly large magnitude of $\gamma_H$ in both samples is
unlikely to arise from coupling of electron transport to
uncorrelated defect dynamics. Long range correlations, such as
many-body Coulomb interactions in doped
semiconductors~\cite{swastik}, however has been known to result in
large $\gamma_H$. In NiTi thin films such correlations may appear
in the form of a long range elastic potential~\cite{shenoy}, in
which defect dynamics takes place. This may also contribute a
non-Gaussian component in the kinetics of resistance fluctuations.
We have plotted the probability distribution function of the
fluctuations in the transition region (170~K $-$ 290~K), in
Fig.~3c. In spite of relatively large device dimensions
(1$\times$10$^{17}$ atoms), existence of a weak non-Gaussian tail
indeed suggests a correlated dynamics of defects. Note that the
premartensitic fluctuations~\cite{reche} or the displacive motion
of twin boundaries in NiTi are primarily athermal, and hence not
expected to have a significant effect at long time scales.

In conclusion, we have investigated the equilibrium low-frequency
resistivity fluctuations, or noise, in the thin films of NiTi
shape memory alloys. The noise magnitude was found to be unusually
large, and displays signature of the martensite transformation on
the dynamics of structural defects. Our investigations suggest
independent estimates of the martensite transformation temperature
scales from a direct microscopic mechanism that agree with those
obtained from resistivity and other structural characterizations.

\textbf{Acknowledgement} \linebreak C. U. and A. G. thank S. Kar
for the STM measurements.

\end{document}